\documentclass[12pt,epsfig]{article}

\usepackage{graphicx}

\usepackage{amssymb}
\usepackage{amsmath}
\usepackage{amscd}
\usepackage{latexsym}

\newcommand{\be}{\begin{equation}}
\newcommand{\ee}{\end{equation}}

\newcommand{\dlt}{\delta}

\newcommand{\ra}{\rightarrow}

\newcommand{\cL}{{\cal L}}
\newcommand{\cM}{{\cal M}}
\newcommand{\cA}{{\cal A}}
\newcommand{\cC}{{\cal C}}
\newcommand{\lgl}{\langle}
\newcommand{\rgl}{\rangle}

\begin{document}

\begin{center}

{\Large {\bf Scheme of thinking quantum systems} \\ [5mm]

V.I. Yukalov$^{1,2}$ and D. Sornette$^{2,3}$ } \\ [3mm]

{\it 
$^1$Bogolubov Laboratory of Theoretical Physics, \\
Joint Institute for Nuclear Research, Dubna 141980, 
Russia, \\ [2mm]

$^2$Department of Management, Technology and Economics, \\
ETH Z\"urich, Swiss Federal Institute of Technology, \\
Kreuzplatz 5, Z\"urich CH-8032, Switzerland, \\ [2mm]

$^3$Swiss Finance Institute, \\
c/o University of Geneva, CH-1211 Geneva 4, Switzerland}

\end{center}

\vskip 1cm

\begin{abstract}
A general approach describing quantum decision 
procedures is developed. The approach can be applied to 
quantum information processing, quantum computing, creation 
of artificial quantum intelligence, as well as to analyzing 
decision processes of human decision makers. Our basic point 
is to consider an active quantum system possessing its own 
strategic state. Processing information by such a system 
is analogous to the cognitive processes associated to 
decision making by humans. The algebra of probability 
operators, associated with the possible options available 
to the decision maker, plays the role of the algebra of 
observables in quantum theory of measurements. A scheme is 
advanced for a practical realization of decision procedures 
by thinking quantum systems. Such thinking quantum systems 
can be realized  by using spin lattices, systems of magnetic 
molecules, cold atoms trapped in optical lattices, ensembles 
of quantum dots, or multilevel atomic systems interacting 
with electromagnetic field. 

\end{abstract}

\vskip 2cm


\vskip 2cm


\newpage

\section{Introduction}

The study of different processes realized by and with quantum 
systems is of great importance for many problems, from the well
established quantum measurements [1] to the currently developing 
fields of quantum information processing and quantum computation 
[2-7]. In all these cases, one considers procedures accomplished 
by an external observer over a {\it passive} quantum system. The 
principal question we address in the present paper is whether a 
quantum system could be {\it active}, and take decisions in a way 
similar to decision making performed by human beings. That is, 
could a quantum system think, as humans do? Answering this question 
is vital for understanding whether a quantum artificial intelligence 
could be created.

It is necessary to stress that the description of human 
thinking processes we have in mind does not refer to the 
physiological mechanisms occurring in the brain, but rather concerns 
the general mathematical scheme that may be developed for 
formalizing such processes. Anticipating on our main theme, if 
the human thinking process could be formalized by a quantum 
mechanics based approach, it would then be straightforward to 
attempt realizing a scheme of such a process by using a suitable 
quantum system.

In order to answer the above question, it is necessary, 
first of all, to assess whether it is possible to describe the 
process of human thinking in terms of the quantum mechanical 
language. Actually, Bohr [8] conjectured that the processes 
involved in human thinking are similar to quantum operations, 
and hence could be described by the language of quantum theory. 
This is in contrast with the standard way of characterizing 
the process of human thinking, which is well formalized by
classical decision theory, whose mathematical foundation is 
due to von Neumann and Morgenstern [9]. This classical approach 
explains the process of human decision making as being based on 
the evaluation of expected utility. Despite its normative appeal, 
the classical expected utility theory is confronted with a number 
of paradoxes, when compared with decision making of real 
human beings [10].      

If we were to follow the Bohr's idea [8] that the human 
thinking processes can be described by means of quantum theory, 
we should stress that this possibility does not require 
the assumption that humans are some quantum systems. Instead, 
it holds that the process of human thinking can be mathematically 
formalized in the language of quantum theory, similarly to the 
process of quantum measurement  [11]. To make our point clear, 
consider the analogous situation presented by the theory of 
differential equations, which was initially developed for 
describing the motion of planets. The theory of differential 
equations is now employed everywhere, being just an efficient 
mathematical tool. In the same way, we explore here the possibility 
that quantum theory may provide a convenient framework for the 
mathematical description of thinking processes.    

The aim of the present paper is twofold. First, we give a general 
mathematical scheme of how decision making can be described in the 
language of quantum theory. The scheme is general in the sense that 
it can be equally applied to human decision makers and to 
active quantum systems imitating the process of taking decisions. 
Second, we suggest the principal architecture for such 
thinking quantum systems and discuss how it could be realized by real 
physical systems, such as ensembles of spins, atoms, molecules, 
or quantum dots.

\section{Mathematical Scheme}

To be precise, we first formulate the mathematical scheme 
characterizing the process of decision making in the 
language of quantum theory.

\vskip 3mm
{\it Definition 1. Action ring}

The process of taking decisions implies that one is deliberating 
between several admissible actions with different outcomes, 
in order to decide which of the intended actions to choose. 
Therefore, the first element arising in decision theory is an 
{\it intended action} $A$. The set of these admissible intended 
actions can be enumerated with an index $i = 1,2,..., N$, where 
the total number $N$ of actions can be finite or infinite. The 
whole family of all these actions forms  the {\it action set}
\be
\label{1}
\cA \equiv \{ A_i : \; i=1,2,\ldots , N \} \; .
\ee
The elements of this set are assumed to be endowed with two binary 
operations, addition and multiplication, so that, if $A$ and 
$B$ pertain to ${\cal A}$, then $AB$ and $A+B$ also pertain 
to ${\cal A}$. The addition is associative, such that 
$A+(B+C)=(A+B)+C$, and reversible, in the sense that $A+B=C$
implies $A = C-B$. The multiplication is distributive, 
$A(B+C)=AB+BC$, and idempotent, $AA = A$. The latter means that 
an intended action, being thought of twice, is the same action. 
The multiplication is not necessarily commutative, so that, 
generally, $AB$ is not the same as $BA$. Among the elements 
of the action set (\ref{1}), there is an identical action $1$, 
for which $A1 = 1A = A$. And there exists an impossible action $0$, 
for which $A0 = 0A = 0$. Two actions are called disjoint, when 
there joint action is impossible, giving $AB = BA = 0$. The action 
set (\ref{1}), with the described structure, is termed the 
{\it action ring}.

\vskip 3mm
{\it Definition 2. Action modes}

\vskip 2mm
An action is simple, when it cannot be decomposed into the 
sum of other actions. An action is composite, when it can be 
represented as a sum of several other actions. If an action is 
represented as a sum
\be
\label{2}
A_i = \bigcup_{\mu=1}^{M_i} A_{i\mu}   \; ,
\ee
whose terms are mutually incompatible, then these terms are 
named the {\it action modes}. The modes correspond to different 
possible ways of realizing an action.

\vskip 3mm
{\it Definition 3. Elementary prospects}

\vskip 2mm
Generally, decision taking is not necessarily associated with 
a choice of just one action among several simple given actions, 
but it involves a choice between several complex actions. The 
simplest such complex action is defined as follows. Let the 
multi-index $n = \{\nu_1, \nu_2,...,\nu_N\}$ be a set of 
indices enumerating several chosen modes, under the condition 
that each action is represented by one of its modes. The 
{\it elementary prospect} is the conjunction
\be
\label{3}
e_n \equiv \bigcap_{i=1}^N A_{i\nu_i}   \; ,
\ee
of the chosen modes, one for each of the actions from the action 
ring (\ref{1}). The total set of all elementary prospects will 
be denoted as $\{e_n\}$. 

\vskip 3mm
{\it Definition 4. Composite prospects}

\vskip 2mm
A prospect is composite, when it cannot be represented as an 
elementary prospect (\ref{3}). Generally, a composite prospect 
is a conjunction 
\be
\label{4}
\pi_j = \bigcap_n A_{j_n}
\ee
of several composite actions of form (\ref{2}), where each of 
the factors $A_{j_n}$ pertains to the action ring (\ref{1}). 

\vskip 3mm
{\it Definition 5. Prospect lattice}

\vskip 2mm
All possible prospects, among which one needs to make a choice, 
form a set
\be
\label{5}
\cL = \{ \pi_j : \; j=1,2,\ldots,N_L \} \; .
\ee
The set is assumed to be equipped with the binary relations 
$ >, <, =, \geq, \leq$, so that each two prospects $\pi_i$ and 
$\pi_j$ in $\cal L$ are related as either  $\pi_i>\pi_j$, or 
$\pi_i=\pi_j$, or $\pi_i\geq\pi_j$, or $\pi_i< \pi_j$, or 
$\pi_i\leq \pi_j$. For a while, it is sufficient to assume that 
such an ordering exists. Then, the ordered set (\ref{5}) is called 
a {\it lattice}. The explicit ordering procedure associated with 
decision making will be given below.      
 
\vskip 3mm
{\it Definition 6. Mode space}

\vskip 2mm
To each action mode $A_{i\mu}$, there corresponds the {\it mode 
state} $|A_{i\mu}\rgl$, which is a complex function $\cA\ra\cC$, 
and its Hermitian conjugate $\lgl A_{i\mu}|$. Here we employ the 
Dirac notation [12]. We assume that a scalar product is defined, 
such that the mode states, pertaining to the same action, are 
orthonormalized:
\be
\label{6}
\lgl A_{i\mu} | A_{i\nu} \rgl = \dlt_{\mu\nu} \; .
\ee
The {\it mode space} is the closed linear envelope
\be
\label{7}
\cM_i \equiv {\rm Span} \{ | A_{i\mu}\rgl : \; 
\mu = 1,2, \ldots, M_i\} \; ,
\ee
spanning all mode states. By this definition, the mode space, 
corresponding to an $i$-action $A_i$, is a Hilbert space of 
dimensionality $M_i$.

\vskip 3mm
{\it Definition 7. Mind space}

\vskip 3mm
To each elementary prospect $e_n$ there corresponds the {\it basic 
state} $|e_n\rgl$, which is a complex function $\cA^N\ra\cC$, and 
its Hermitian conjugate $\lgl e_n|$. The structure of a basic state 
is
\be
\label{8}
| e_n\rgl \equiv | A_{1\nu_1} A_{2\nu_2} \ldots 
A_{N\nu_N} \rgl = \bigotimes_{i=1}^N | A_{i\nu_i}\rgl  \; . 
\ee
The scalar product is assumed to be defined, such that the basic 
states are orthonormalized:
\be
\label{9}
\lgl e_m | e_n \rgl = \prod_{i=1}^N \dlt_{\mu_i\nu_i}
\equiv \dlt_{mn} \; .
\ee
The {\it mind space} is the closed linear envelope
\be
\label{10}
\cM \equiv {\rm Span} \{ | e_n\rgl \} = 
\bigotimes_{i=1}^N \cM_i \;  ,
\ee
spanning all basic states (\ref{8}). Hence, the mind space is a 
Hilbert space of dimensionality
$$
{\rm dim} \cM = \prod_{i=1}^N M_i \; .
$$
The vectors of the mind space represent all possible actions and 
prospects considered by a decision maker.

\vskip 3mm
{\it Definition 8. Prospect states}

\vskip 2mm
To each prospect $\pi_j$ there corresponds a state $|\pi_j\rgl\in\cM$ 
that is a member of the mind space (\ref{10}). Hence, the prospect 
state can be represented as an expansion over the basic states
\be
\label{11}
|\pi_j \rgl = \sum_n \; a_{jn} | e_n \rgl \; .
\ee
The prospect states are not required to be mutually orthogonal and 
normalized to one, so that the scalar product
$$
\lgl \pi_i | \pi_j \rgl = \sum_n \; a_{in}^* a_{jn}
$$
is not necessarily a Kronecker delta.

\vskip 3mm
{\it Definition 9. Strategic state}

\vskip 2mm
Among the states of the mind space, there exists a special fixed 
state $|s\rgl\in\cM$, playing the role of a reference state, which 
is termed the {\it strategic state}. This is the state characterizing 
the specific decision maker. Being in the mind space (\ref{10}), this 
state can be represented as the decomposition
\be
\label{12}
| s \rgl = \sum_n c_n | e_n \rgl \; .
\ee
Being a unique state, characterizing each decision maker
like its fingerprints, it can be normalized 
to one:
\be
\label{13}
\lgl s | s \rgl = 1 \; .
\ee
From Eqs. (\ref{12}) and (\ref{13}), it follows that 
$$
\sum_n | c_n|^2 = 1   .
$$
The existence of the strategic state, uniquely defining each 
particular decision maker, is the principal point distinguishing 
the active thinking quantum system from a passive quantum system 
subject to measurements from an external observer. For a passive 
quantum system, predictions of the outcome of measurements are 
performed by summing (averaging) over all possible statistically 
equivalent states, which can be referred to as a kind of 
``annealed''  situation. In contrast, decisions and observations 
associated with a thinking quantum system occur in the presence 
of this unique strategic space, which can be thought of as a kind 
of fixed ``quenched'' state. As a consequence, the outcomes of the 
applications of the quantum mechanical formalism will thus be 
different for thinking versus passive quantum systems.

\vskip 3mm
{\it Definition 10. Prospect operators}

\vskip 2mm
Each prospect state $|\pi_j\rgl$, together with its Hermitian 
conjugate $\lgl\pi_j|$, defines the {\it prospect operator}
\be
\label{14}
\hat P(\pi_j) \equiv | \pi_j \rgl \lgl \pi_j | \; .
\ee
By this definition, the prospect operator is self-adjoint. The 
family of all prospect operators forms the involutive bijective 
algebra that is analogous to the algebra of local observables in 
quantum theory. Since the prospect states, in general, are neither 
mutually orthogonal nor normalized, the squared operator
$$
\hat P^2(\pi_j) = \lgl \pi_j | \pi_j \rgl \hat P(\pi_j)
$$  
contains the scalar product
$$
 \lgl \pi_j | \pi_j \rgl = \sum_n | a_{nj}|^2 \;  ,
$$
which does not equal to one. This tells us that the prospect 
operators, generally, are not idempotent, thus, they are not 
projection operators. It is only when the prospect is elementary 
that the related prospect operator 
$$
\hat P(e_n) = | e_n \rgl \lgl e_n |
$$
becomes idempotent and is a projection operator. But, in general, 
this is not so.

\vskip 3mm
{\it Definition 11. Prospect probabilities}

\vskip 2mm
In quantum theory, the averages over the system state, for 
the operators from the algebra of local observables, define the 
observable quantities. In the same way, the averages, over the 
strategic state, for the prospect operators define the observable 
quantities, the {\it prospect probabilities}
\be
\label{15}
p(\pi_j) \equiv \lgl s | \hat P(\pi_j) | s \rgl  \; .
\ee
These are assumed to be normalized to one:
\be
\label{16}
\sum_{j=1}^{N_L}\; p(\pi_j) = 1 \; ,
\ee
where the summation is over all prospects from the prospect lattice 
(\ref{5}). By their definition, quantities (\ref{15}) are non-negative, 
since Eq. (\ref{15}) reduces to the modulus of the transition 
amplitude squared
$$
p(\pi_j) = | \lgl \pi_j | s \rgl |^2  \; .
$$
Being normalized as in Eq. (\ref{16}), the set $\{p(\pi_j)\}$ composes 
the scalar probability measure.

\vskip 3mm
{\it Definition 12. Utility factor}

\vskip 2mm
The diagonal form
\be
\label{17}
p_0(\pi_j) \equiv \sum_n \;
\lgl s | \hat P(e_n) \hat P(\pi_j) \hat P(e_n) | s \rgl
\ee
plays the role of the expected utility in classical decision making, 
justifying its name as the {\it utility factor}. In order to 
be generally defined and to be independent of the chosen units of measurement, 
the factor (\ref{17}) can be normalized as
\be
\label{18}
\sum_{j=1}^{N_L} \; p_0(\pi_j) = 1 \; .
\ee
The fact that factor (\ref{17}) is really equivalent to the classical 
expected utility follows from noticing that
$$
\hat P(e_n) | s \rgl = c_n | s \rgl \; ,
$$  
hence Eq. (\ref{17}) acquires the form
$$
p_0(\pi_j) = \sum_n | c_n|^2 \lgl e_n | \hat P(\pi_j) | e_n \rgl \; ,
$$
where $<e_n|\hat{P}(\pi_j)|e_n>$ plays the role of a utility function, 
weighted with the probability $|c_n|^2$.

\vskip 3mm
{\it Definition 13. Attraction factor}

\vskip 2mm
The nondiagonal term
\be
\label{19}
q(\pi_j) \equiv \sum_{m\neq n} \;
\lgl s | \hat P(e_m) \hat P(\pi_j) \hat P(e_n) | s \rgl
\ee
arises as a consequence of the quantum interference effect. Its 
appearance is typical of quantum mechanics. Such nondiagonal terms 
do not occur in classical decision theory. This term can be called 
the {\it interference factor}. Interpreting its meaning in decision 
making, we can associate its appearance as resulting from the system 
deliberation between several alternatives, when deciding which 
of the latter is more attractive. Thence, the name ``attraction 
factor.'' Using expansion (\ref{12}) in Eq. (\ref{19}) yields
$$
q(\pi_j) \equiv \sum_{m\neq n} c_m^* c_n \lgl e_m |
\hat P(\pi_j) | e_n \rgl \; ,
$$
which shows that the interference occurs between different 
elementary prospects in the process of considering a composite 
prospect $\pi_j$. It is worth stressing that the interference 
factor is nonzero only when the prospect $\pi_j$ is composite. 
If it were elementary, say $\pi_j=e_k$ then, since
$$
\hat P(e_k) | e_n \rgl = \dlt_{nk} | e_n \rgl \; ,
$$
we would have
$$
q(e_k) = \sum_{m\neq n} c_m^* c_n \dlt_{mn}
\dlt_{nk} = 0 \; ,
$$
and no interference would arise.

\vskip 3mm  
{\it Definition 14. Prospect ordering}

\vskip 2mm
In defining the prospect lattice (\ref{5}), we have assumed 
that the prospects could be ordered. Now, after introducing 
the scalar probability measure, we are in a position to give 
an explicit prescription for the prospect ordering. We say that 
the prospect $\pi_1$ is {\it preferable} to $\pi_2$
if and only if  
\be
\label{20}
p(\pi_1) > p(\pi_2) \qquad ( \pi_1 > \pi_2) \;  .  
\ee
Two prospects are called indifferent if and only if
\be
\label{21}
p(\pi_1) = p(\pi_2) \qquad ( \pi_1 = \pi_2) \; .
\ee
And the prospect $\pi_1$ is preferable or indifferent to $\pi_2$ 
if and only if
\be
\label{22}
p(\pi_1) \geq p(\pi_2) \qquad ( \pi_1 \geq \pi_2) \; .
\ee
These binary relations provide us with an explicit prospect ordering 
making the prospect set (\ref{5}) a lattice.

\vskip 3mm
{\it Definition 15. Optimal prospect}

\vskip 2mm
Since all prospects in the lattice are ordered, it is straightforward 
to find among them that one enjoying the largest probability. This 
defines the {\it optimal prospect} $\pi_*$ for which
\be
\label{23}
p(\pi_*) \equiv \sup_j p(\pi_j) \; .
\ee
Finding the optimal prospect is the final goal of the decision-making 
process. Since the prospect probabilities are non-negative, it is 
possible to find the minimal prospect  in the lattice (\ref{5}) with 
the smallest probability. And the largest probability defines the 
optimal prospect $\pi_*$. Therefore the prospect set (\ref{5}) is 
a complete lattice.

From the above definitions, we get the following theorem.

\vskip 2mm
{\it Theorem}. The prospect probability (\ref{15}) has the form
\be
\label{24}
p(\pi_j) = p_0(\pi_j) + q(\pi_j) \;  ,
\ee
in which the attraction factor (\ref{19}) satisfies the {\it 
alternation property}
\be
\label{25}
\sum_{j=1}^{N_L} \; q(\pi_j) = 0 \;  ,
\ee
where the summation is over the whole lattice $\cal L$.

\vskip 2mm
{\it Proof}. Substituting into definition (\ref{15}) expansion 
(\ref{12}) and invoking Eqs. (\ref{17}) and (\ref{19}) gives 
expression (\ref{24}). Employing the normalization conditions (\ref{16}) 
and (\ref{18}) results in the alternation property (\ref{25}).  

\vskip 2mm
The above definitions and the theorem constitute the basic 
mathematical structure of the Quantum Decision Theory (QDT), which 
can be applied to different decision makers, whether these are human 
beings or quantum systems.

\section{Physical Scheme}

{\parindent=0pt
{\bf 3.1. General scheme of a thinking quantum machine}

\vskip 3mm

The basic elements of a thinking quantum machine are as follows. 
As is evident, no decision maker can be in equilibrium and be 
a closed system. Therefore, there should be an external energy 
supply supporting the strategic state characterizing the 
particular decision maker. The prospect states are also imposed 
from outside, representing the information loaded into the system, 
about which a decision is to be taken. Measuring the averages of 
the prospect operators is equivalent to measuring the averages 
of observable quantities. In the present case, the prospect states 
are scattered over the strategic state. The measurement procedure 
results in the scattering cross-sections that are proportional to
the scattering amplitudes squared. The latter define the prospect 
probabilities. The prospect analyzer selects the largest probability, 
and in so doing chooses the optimal prospect. The state corresponding 
to the optimal prospect is the system output. This summarizes how 
the system processes information, imitating the procedure of human
decision making, as shown in Fig. 1.        

\vskip 5mm

{\bf 3.2. Physical embodiment of thinking quantum machines}

\vskip 5mm

Let us now suggest different physical quantum systems that could 
realize the process of decision making based on the mathematical 
theory of Sec. 2 and on the general scheme of the previous subsection 
illustrated by Fig. 1. Several physical systems can be used for this 
purpose. The main requirement is that the system would be composed of 
several parts, enumerated with an index $i = 1,2,...,N$, and that, in 
each $i$-part, mode states $|A_{i\mu}\rgl$ could be generated. Such 
systems can be realized with real-space lattices or nanostructures. 
Below, we discuss several possible physical realizations.  

\vskip 3mm
(i) {\it Real-space lattice of  spins}

\vskip 2mm
Such a very convenient system is the real-space lattice of 
spins. Then the mode state $|A_{i\mu}\rgl$ is the spin state in 
the $i$-lattice site, with the $\mu$-spin projection. The basic 
states $|e_n\rgl$ are the lattice spin states forming the basis 
for the given lattice. The prospect states $|\pi_j\rgl$ are the 
states produced by external fields that play the role of information 
provided to the system. The strategic state $|s\rgl$ is a chosen 
state supported by means of additional energy supply. Different 
nontrivial many-body spin states can be created by applying, for 
instance, staggered magnetic fields to a spin lattice [13]. In some 
cases [14], the staggered magnetization arises in spin systems by 
its own. The spins in a lattice can be of different physical nature 
and of different type. Promising candidates could be magnetic 
molecules having high spins, hence providing a number of modes 
in each lattice site (see details in review articles [15-18]). 
The manipulation with magnetic molecules can be realized by 
swiping magnetic fields. An ultrafast manipulation with lattice 
spins can be done by connecting the sample to a resonator, which 
makes it possible to create various regimes of spin motion [17-25]. 
Magnetic clusters of high spins [26,27] can be used as well. Their 
spins can also be manipulated by external magnetic fields, achieving 
a fast regulation in the presence of a resonator [17,18,23-25]. 

\vskip 3mm
(ii) {\it Optical lattices of cold Bose-condensed atoms}

\vskip 2mm
Optical lattices of cold Bose-condensed atoms could be another 
physical system allowing one to create various atomic 
states [28-33]. When each lattice site contains many condensed 
bosons, it is possible [34-36] to realize a resonant generation 
of coherent topological modes in each of the lattice sites. Then 
the mode state $|A_{i\mu}\rgl$ is a $\mu$-type coherent mode in 
the lattice $i$-site. The manipulation of the modes can be 
accomplished by means of a magnetic field modulation for atoms 
trapped in the lattice sites [34-36] or by modulating the atomic 
scattering length by means of Feshbach resonance techniques [37]. 
The overall setup is similar to that of the spin systems.

\vskip 3mm 
(iii) {\it Bosons or fermions in double-well optical lattices}

\vskip 2mm
In the case of double-well optical lattices [38-42], one can use 
bosons as well as fermions by creating different insulating states 
related to atoms shifted into one or another well of a double well 
[43-45], with the index $\mu$ marking the left or right side of 
a double well. Then the mode state $|A_{i\mu}\rgl$ corresponds to 
the left or right position of an atom in a double well located at 
the lattice $i$-site. By varying the parameters of a double well, 
different atomic states can be created [43-46]. 

\vskip 3mm
(iv) {\it Multilevel radiating atoms or molecules inside a 
solid-state matrix}

\vskip 2mm
Multilevel radiating atoms or molecules inside a solid-state 
matrix can also be used, provided it is feasible to transfer each 
of the atoms into the required energy states by applying external 
laser beams. The state $|A_{i\mu}\rgl$ is then the state of an 
$i$-atom on the $\mu$-level. Atomic interactions through the 
common radiation field would result in highly correlated coherent 
states of the whole system [47,48]. Multilevel atoms seem to be a 
convenient tool for realizing quantum computation and quantum 
information processing [49,50]. 

\vskip 3mm
(v) {\it Nanostructures}

\vskip 2mm
Nanostructures, such as quantum dots, quantum wells, and quantum 
wires, possessing discrete energy levels, are known [51-54] to 
have properties similar to multilevel atoms, justifying their 
name ``artificial atoms". Therefore, the assemblies of such 
nano-objects can also be used in the same way as atomic or 
molecular systems. Having sizes essentially larger than atoms, 
these quantum nanostructures make it easier to regulate the mode 
states of each separate object. At the same time, they can interact 
with each other forming a common many-body state for the whole 
sample [55].           
}

\section{Conclusion}

In conclusion, we have described a mathematical approach 
characterizing the process of decision making in the language 
of quantum theory. We have suggested a general physical scheme 
realizing this process by an active quantum system, imitating the 
procedure of decision making. Being an active system, it mimics
the process of thinking, because of which such systems can be called
{\it thinking quantum systems}.

We have proposed  several physical substances which could 
be used for implementing the desired architecture for a thinking 
quantum system that would imitate the process of decision making. 
The technicalities for the practical construction of such systems 
would, of course, depend on the chosen physical matter. But it 
is not our goal to discuss the technical details, which is rather 
the privilege of experimentalists. 

The developed approach can be used for quantum information processing, 
quantum computing, and for creating artificial quantum intelligence.

\vskip 5mm
{\bf Acknowledgements}

\vskip 3mm
We are grateful to E.P. Yukalova for useful discussions. One of 
the authors (V.I.Y.) appreciates a grant from the Russian Foundation 
for Basic Research.  

\newpage

\begin{figure}[h]
\centerline{\includegraphics[width=16.5cm]{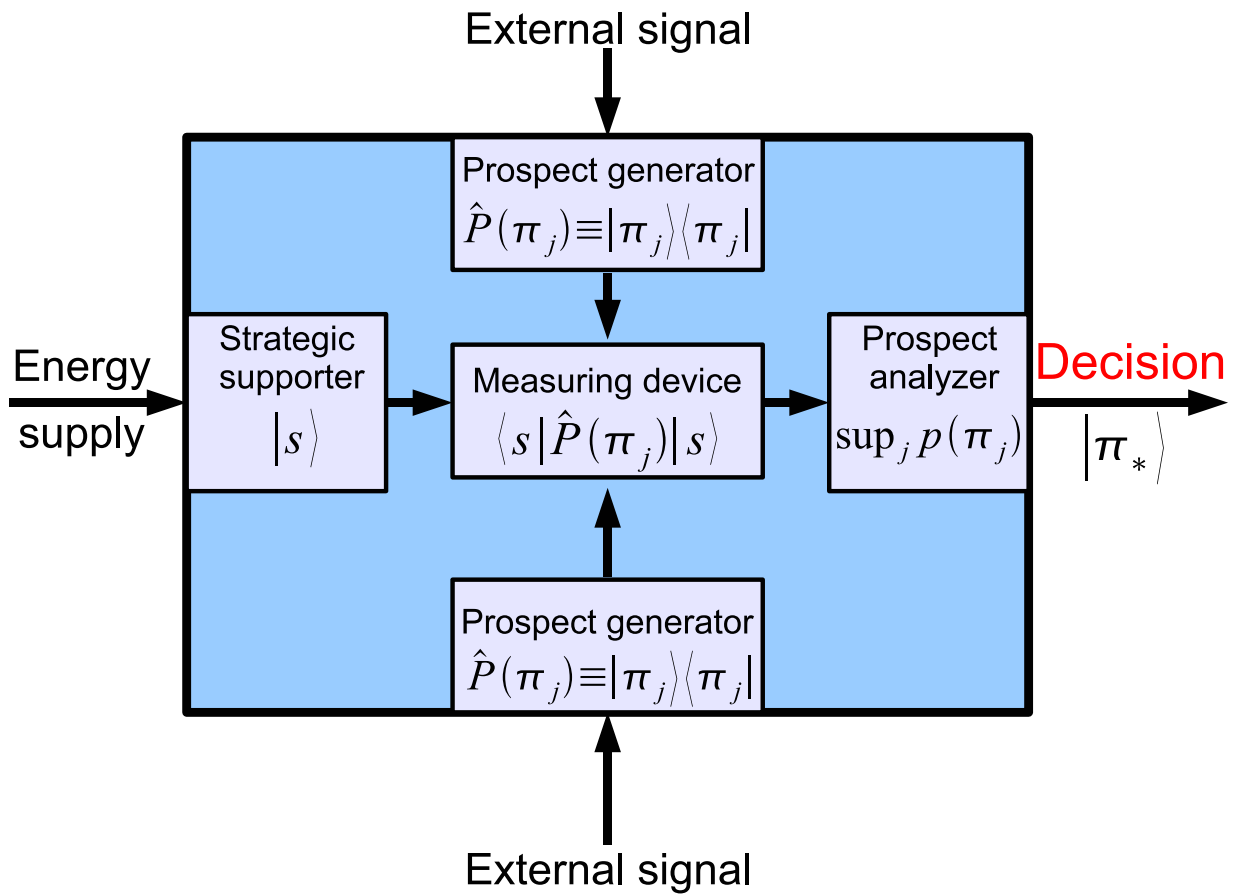}}
\caption{Principal scheme of a thinking quantum system, imitating 
the process of decision making.} 
\label{fig:Fig.1}
\end{figure}

\end{document}